# A System-View Optimal Additional Active Power Control of Wind Turbines for Grid Frequency Support

Yubo Zhang, *Student Member, IEEE*, Zhiguo Hao, and Songhao Yang, *Member, IEEE*, Baohui Zhang, *Fellow, IEEE*

*Abstract*—Additional active power control (AAPC) of wind turbines (WTs) is essential to improve the transient frequency stability of low-inertia power systems. Most of the existing research has focused on imitating the frequency response of the synchronous generator (SG), known as virtual inertia control (VIC), but are such control laws optimal for the power systems? Inspired by this question, this paper proposes an optimal AAPC of WTs to maximize the frequency nadir post a major power deficit. By decoupling the WT response and the frequency dynamics, the optimal frequency trajectory is solved based on the trajectory model, and its universality is strictly proven. Then the optimal AAPC of WTs is constructed reversely based on the average system frequency (ASF) model with the optimal frequency trajectory as the desired control results. The proposed method can significantly improve the system frequency nadir. Meanwhile, the event insensitivity makes it can be deployed based on the on-line rolling update under a hypothetic disturbance, avoiding the heavy post-event computational burden. Finally, simulation results in a two-machine power system and the IEEE 39 bus power system verify the effectiveness of the optimal AAPC of WTs.

## I. Introduction

WIND power plays an indispensable role in the energy transition and carbon neutral for its clean and efficient [1]. As the mainstream form of wind energy utilization, the large-scale installation of the grid-following type of wind turbines (WTs) significantly impairs the inertia of power systems. The low inertia trend implies weak resistance to power disturbances, which raises a serious concern for the frequency stability of power systems.

To improve the frequency stability of low-inertia power systems, the WT is requested to provide proper frequency support. Fortunately, vector control allows its output power to be rapidly and precisely adjusted within a reasonable range. With this feature, the additional active power control (AAPC) is enabled to directly regulate the WT active power in response to the frequency excursion. In general, the existing AAPC can be roughly divided into three categories, namely the WT-view, the wind farm (WF)-view, and the system-view methods.

The WT-view AAPC is fully decentralized, intending to excavate the frequency support ability of the WT itself. The most popular WT-view AAPC method is the virtual-inertia control (VIC). The WT with VIC generates an additional active power that depends on the frequency excursion, the rate of change of frequency (RoCoF), or both of them. The VIC with fixed parameters was first proposed in [2], and this pioneering work revealed the potential of the WT in the grid frequency support. Considering the limitation of fixed parameters to varying scenarios, adaptive VIC attracts extensive attention [3-6]. In [3] and [4], the proportional-derivative coefficients of the VIC were adaptively adjusted according to the WT's releasable kinetic energy. Using the input-output feedback linearization method, a time-varying controller was designed to provide an inertial response in [5]. Besides, an adaptive droop control of the WT was proposed [6]. In [7], both the RoCoF and the WT's rotor speed were taken into consideration, aiming to mitigate the degradation of the VIC under severe contingencies. To avoid the strong dependence on the measured frequency and VIC coefficients, a fast frequency regulation (FR) strategy of the WT based on variable power point tracking control was proposed in [8]. Overall, the WT-view AAPC of the WT is self-governed, simple and highly reliable, but lacks collaboration with other WTs and perception of system demands.

Inspired by optimal control such as model predictive control (MPC), the WF-view AAPC strategies have been extensively studied on the cooperation of WTs. In [9], the WF-view power reference was properly shared among WTs by solving an MPC problem, of which the optimization goal is minimizing wind energy loss, etc. Further, the cooperation of multiple WFs was included in the optimization model in [10]. To deal with the inaccuracy of MPC prediction, a double-layer control framework was proposed to adequately exploit the kinetic energy of all WTs [11]. In [12], the data-driven MPC method was proposed to cope with the challenge of complex and nonlinear WT dynamics. In the aspect of improving the computational efficiency, the distributed Newton method endowed the proposed method with a super-linear convergence rate in [13]; besides, the off-line look-up table was mentioned in [14] and WTs clustering simplification was proposed in [15].

The system-view strategies expect to explore the feasible AAPC of WTs from the perspective of system frequency optimization. For example, the particle swarm algorithm was applied in [16] to derive the optimal FR control parameter of the WT with the goal of improving the frequency nadir as well as the average frequency. However, the optimization results were event-sensitive to some extent. In [17], a multi-layer MPC was proposed to achieve both objectives of dynamically optimal power support of the WF and the stability of WTs. Despite the improved performance, MPC may unduly consume

This work was supported by National Natural Science Foundation of China (52007143), China Postdoctoral Science Foundation (2021M692526) and Key Research and Development Program of Shaanxi(2022GXLH-01-06).
Y. Zhang, Z. Hao, S. Yang and B. Zhang are with State Key Laboratory of Electrical Insulation and Power Equipment, Xi'an Jiaotong University, Xi'an, China (e-mail: zyb970305@stu.xjtu.edu.cn, {zhghao, songhaoyang, bhzhang}@xjtu.edu.cn).

the power support capacity of WTs in a short time. An optimal synthetic inertia control considering the variation of mechanical power of the WT is proposed in [18], which avoids the secondary drop of system frequency while satisfying the preset frequency nadir constraint. To minimize the RoCoF of power systems, ref. [19] presented a synthetic inertia controller for the WT based on the $H_2$ optimal control method. However, the mechanical power of SGs was required to be measured in real-time, which is difficult in practice.

To further exploit the potential of WTs on improving system frequency stability, this paper proposes a system-view optimal AAPC of WTs. The main contributions of this work include:

1) The optimal frequency trajectory that maximizes the frequency nadir is given and its universality is strictly proved. By decoupling the WT response and the frequency dynamics, the optimal frequency trajectory is efficiently solved by the trajectory optimization model. Furthermore, the universality of the optimal frequency trajectory is strictly proved from the perspective of energy, regardless of the system parameters and the disturbance magnitudes.
2) The feedback-form optimal AAPC of WTs is derived from the ASF model to achieve the optimal frequency trajectory. In addition, an exit strategy is designed as a supplement to the optimal AAPC to ensure the complete state recovery of the WT.
3) The proposed method possesses the following two main advantages. In terms of control effects, the optimal AAPC of WTs maximizes the system frequency nadir. It adaptively responds to power system demands rather than blindly grabbing the WT's frequency support capability. In terms of application, event insensitivity makes the optimal AAPC of WTs can be solved and deployed based on the on-line rolling update under a hypothetic disturbance, which avoids the heavy post-event computational burden.

The rest of the paper is organized as follows. Section II briefly introduces the system frequency and the WT model. Section III shows the design process of the optimal AAPC. The proof of the universality of the optimal AAPC is exhibited in Section IV. Then, the performance of the optimal AAPC is verified in Section V and Section VI. Finally, Section VII draws a conclusion.

## II. SYSTEM FREQUENCY AND WT MODELING

### A. Dynamic Model of Power System Frequency

When the power system suffers a power disturbance, the system frequency will deviate from the nominal value and undergo a transient process. Considering the frequency support of WTs, the frequency dynamic posts a large disturbance can be modeled as the following swing equation

$$2H\frac{d\Delta f(t)}{dt} = \Delta P_m(t) + \Delta P_e(t) - P_d - D\Delta f(t), \quad (1)$$

where $H$ and $D$ denote the normalized inertia constant and damping constant; $\Delta f(t)$ is the frequency excursion; $\Delta P_m$ and $\Delta P_e$ are the FR power of SGs and that of WTs, respectively; $P_d$

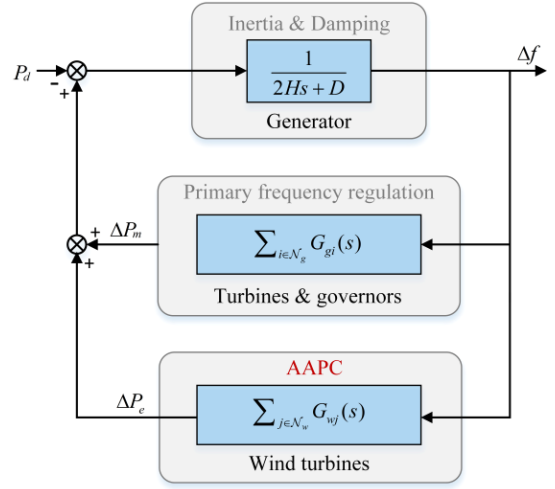

Fig. 1 ASF model considering the AAPC of WTs.

is the power disturbance. Considering that the frequency drop caused by a major power deficit is more common and severe, this paper focuses on the scenario of a frequency drop.

During the primary frequency regulation (PFR), generators' governors and AAPC of WTs actively respond to the power imbalance. This process can be described using an average system frequency (ASF) model [20], as shown in Fig. 1. As for the symbol in Fig. 1, $\mathcal{N}_g$ and $\mathcal{N}_w$ denote the set of traditional SGs and WTs, respectively. $G_{gi}(s)$ is the FR dynamics the governor; $G_{wj}(s)$ is the AAPC of the WT. The subscript "$i$" denotes the $i$-th SG, and the subscript "$j$" denotes the $j$-th WT, the same as below.

In this paper, WTs are supposed to operate in the maximum power point tracking (MPPT) mode and only provide short-term frequency support. Hence, the steady-state frequency excursion of the PFR is not affected. According to [21], it can be calculated as

$$\Delta f_{ss}^* = -\frac{P_d}{D + K_g} \quad (2)$$

where

$$K_g = \sum_{i\in\mathcal{N}_g} P_{gi}/(R_{gi}S_b)$$

and $P_{gi}$ is the rated power of the SG; $1/R_{gi}$ is the steady-state gain factor of the governor of the SG.

Nevertheless, the transient performance of PFR can be effectively improved by optimizing the AAPC of WTs, which has been confirmed in extensive research. As one of the most significant indicators of PFR, the frequency nadir reflects the worst transient frequency drop post a power deficit. Therefore, this paper focuses on exploring the optimal AAPC to maximize the frequency nadir.

### B. WT model

In this paper, one of the most widely-used variables speed WT is considered, namely the doubly-fed induction generator (DFIG) WT. The turbine power of a WT is calculated by the following model.

$$P_t = 0.5\rho\pi R^2 C_p(\lambda,\beta)v_w^3 \quad (3)$$

where $\rho$ is the air density; $R$ is the turbine radius; $v_w$ is the wind speed; $\lambda = R\omega_r/v_w$ is the tip-speed ratio; $\omega_r$ is the rotor speed; $\beta$ is the pitch angle; $C_p(\lambda,\beta)$ is the power coefficient, which is related to the tip-speed ratio and pitch angle. The widely used turbine efficiency model is adopted in this paper.

$$C_p(\lambda,\beta) = 0.22(\frac{116}{\bar{\lambda}} - 0.4\beta - 5)e^{-\frac{12.5}{\bar{\lambda}}},$$
$$\bar{\lambda} = \frac{1}{\lambda + 0.08\beta} - \frac{0.035}{\beta^3 + 1} \quad (4)$$

The mechanical dynamics of the WT depend on its turbine power and output power. According to [22], the one-mass model is adequate to investigate the WT dynamics during the frequency support, as follows.

$$J\omega_r\dot{\omega}_r = P_t - P_e \quad (5)$$

where $J$ is the inertia of the motion system of the WT.

*C. Constraints in Frequency Support of WTs*

The short-term frequency support of WTs is essentially to inject excess active power into the system at the expense of releasing the kinetic energy of the rotor. Inevitably, this additional active power will cause the speed of WTs to decrease and deviate from the steady state. Hence, the released energy is required to be injected back into WTs for the rotor speed recovery. Ignoring the variation of the turbine power, the energy absorption characteristic of WTs is modeled as follows.

$$\int_{t_0}^{t_f} \Delta P_e(t)dt = 0 \quad (6)$$

where $t_0$ and $t_f$ are the start time and the end time of the frequency support of WTs, respectively. For simplicity, the start time $t_0$ is set to 0. To properly exploit the frequency support potential of WTs as well as ensure their safe operation, the end time of frequency support is recommended as 30s, which corresponds to the time scale of the PFR.

In addition, operating constraints such as the output power and the rotor speed unsurprisingly affect the frequency support performance of WTs. However, these operational constraints are generally event-sensitive. That is to say, whether constraints work or not is determined by the severity of the disturbance. Given the unpredictability of the disturbance, this paper expects to explore a more general and event-insensitive AAPC of WTs, thus avoiding heavy computation. Therefore, these event-sensitive constraints of the WT are ignored when deriving the optimal AAPC, and only the "energy absorption" characteristic is retained. Notably, the operating constraints of the WT are formally ignored in the theoretical derivation, which are completely retained in the simulation model.

## III. OPTIMAL AAPC DESIGN

For the high-order ASF model shown in Fig. 1, it is intractable to directly derive the optimal AAPC of WTs that maximizes the frequency nadir. Hence, this paper explores the optimal AAPC of WTs based on reverse thinking. By decoupling the WTs response and frequency dynamics, the theoretically optimal frequency trajectory is freely explored. Then, the optimal AAPC is reversely constructed from the ASF model with the optimal frequency trajectory as the desired control result.

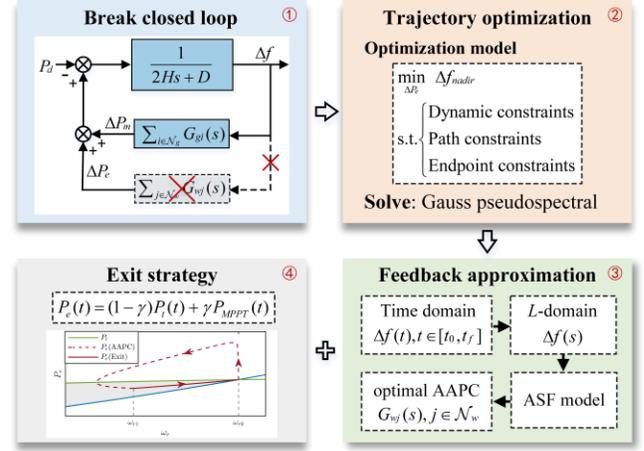

Fig. 2 Framework of the optimal AAPC design.

The specific design framework is shown in Fig. 2. Firstly, the closed loop in the ASF model is broken and an aggregate additional power is introduced for decoupling the calculation of WTs response and frequency dynamics. Then, taking the additional power of WTs as the control variable, a trajectory optimization model is constructed to maximize the frequency nadir. Based on the time-domain solution of the trajectory optimization model, the optimal AAPC of WTs can be reversely constructed from the ASF model. Finally, considering the model error caused by the ignorance of the variation of turbine power of WTs, a proper exit strategy is proposed as a supplement to ensure the speed recovery of WTs.

*A. Breaking the Closed Loop of the ASF Model*

For the high-order ASF model shown in Fig. 1, it is difficult to derive its analytical dynamic in the time domain. Meanwhile, the frequency support of WTs is coupled with the frequency, which causes an obscure impact of AAPC on the frequency nadir. The complex coupling and high-order properties make it challenging to directly solve the optimal AAPC of WTs.

Generally, WTs are considered to be able to output any reasonably desired power curve given their excellent controllability. Therefore, this paper simplifies the problem by breaking the closed loop of the ASF model. Specifically, the additional power of all WTs is aggregated as an independent controllable variable. As an independent control variable, the aggregated additional power of WTs exerts a unidirectional effect on the frequency, which makes it possible to freely explore the optimal control curve of WTs that maximizes the frequency nadir.

*B. Trajectory Optimization Model*

Taking the aggregate additional power of all WTs as the independent control variable, the trajectory optimization model can be constructed to solve the optimal control curve of WTs and the corresponding frequency trajectory in the time domain [23]. Different from the MPC which needs to solve the optimization model in real-time to generate single-step control, the trajectory optimization model derives the optimal control curve in the entire period at once. So, the trajectory optimization model does not require real-time calculation. The construction of the trajectory optimization model is introduced below.

First, a new state variable associated with the released energy of WTs is introduced in (7), aiming to convert the integral constraint shown in (6) into scope constraints.

$$\Delta \dot{E}(t) = \Delta P_e(t), t \in [t_0, t_f] \qquad (7)$$

Then, the integral constraint in (6) can be converted into

$$\Delta E(t_f) - \Delta E(t_0) = 0 \qquad (8)$$

The dynamic constraints consist of the swing equation, governor dynamics, and Eq. (7), which can be expressed as follows.

$$\dot{x}(t) = Ax(t) + B_1 \Delta P_e(t) + B_2 P_d, t \in [t_0, t_f] \qquad (9)$$

where,

$$x(t) = \begin{bmatrix} \Delta f(t) \\ \Delta x_g(t) \\ \Delta E(t) \end{bmatrix}, A = \begin{bmatrix} (D_g - D)/2H & C_g/2H & 0 \\ B_g & A_g & 0 \\ 0 & 0 & 0 \end{bmatrix},$$

$$B_1 = \begin{bmatrix} 1/2H \\ 0 \\ 1 \end{bmatrix}, B_2 = \begin{bmatrix} -1/2H \\ 0 \\ 0 \end{bmatrix},$$

$\Delta P_e(t)$ is the control variable; $A_g$, $B_g$, $C_g$, $D_g$, and $x_g$ model the dynamics of governors, see Appendix A for details.

Since the power system is supposed to operate at the steady-state point initially, the initial values of all the state variables in (9) are 0. Combined with (8), the endpoint constraints of the trajectory optimization model are as follows.

$$\begin{cases} \Delta f(t_0) = \Delta E(t_0) = \Delta E(t_f) = 0 \\ \Delta x_g(t_0) = 0 \end{cases} \qquad (10)$$

A static variable $\Delta f_{nadir}$ is introduced, denoting the deviation of the frequency nadir relative to the base frequency.

$$\Delta f_{nadir} = f_{nadir} - f_B \qquad (11)$$

where $f_{nadir}$ is the frequency nadir in the transient process of PFR; $f_B$ is the base value of the system frequency.

Naturally, the frequency variation should not exceed this value in the transient process.

$$\Delta f(t) \geq \Delta f_{nadir}, t \in [t_0, t_f] \qquad (12)$$

where $\Delta f(t)$ and $\Delta f_{nadir}$ are both negative.

Finally, the following optimization objective is designed from a system perspective, aiming to improve the frequency nadir.

$$\max_{\Delta P_e} \Delta f_{nadir} \qquad (13)$$

Based on the above processing, the following trajectory optimization model can be constructed. Specifically, Eq. (13) is the optimization objective; Eq. (9) is the dynamic constraint; Eq. (12) is the path constraint; Eq. (10) is the endpoint conditions. Then, the optimal control curve and frequency trajectory in the time domain can be solved efficiently by the Gauss pseudospectral method [24], which is briefly introduced in Appendix B.

The solution of the trajectory optimization model corresponding to the two-machine system in Section V is shown in Fig. 3. Specifically, Fig. 3(a) shows the aggregate additional power of WTs, which is positive in the initial stage to suppress the frequency drop, and negative in the later process for rotor speed recovery. Accordingly, the kinetic energy is released initially and then entirely absorbed from the grid, as shown in Fig. 3(b). The frequency dynamics shown in Fig. 3(c) indicate the optimal control of WTs maintains the frequency at a constant nadir, denoted as $\Delta f^*_{nadir}$, even in the process of energy absorption. Fig. 3(d) further shows the trajectory of the equivalent WT. Ignoring the variation of turbine power, WTs theoretically recover to the steady-state operation at the end of frequency support, which is guaranteed by the constraint in (6).

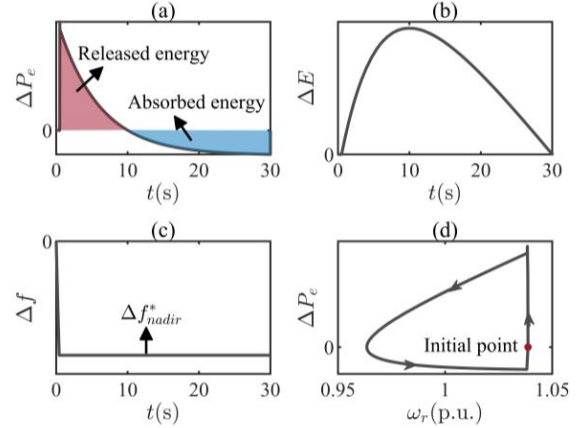

Fig. 3 Time-domain solution of trajectory optimization model. (a) Optimal additional power of WTs. (b) Released energy of WTs. (c) Optimal frequency dynamics. (d) Trajectory of the equivalent WT.

*C. Feedback Approximation*

The optimal frequency trajectory in the time domain is obtained by solving the trajectory optimization model. Then, the optimal feedback-form AAPC of WTs is derived based on reverse approximation, as follows.

By observing the shape of the optimal frequency trajectory in Fig. 3(c), the frequency dynamics can be approximated as the following exponential decay form, as shown in Fig. 4.

$$\Delta f(t) = a(1 - e^{-bt}) \qquad (14)$$

When $t \to \infty$, the frequency should be equal to $\Delta f^*_{nadir}$. So the coefficient $a$ can be derived as follows.

$$a = \Delta f^*_{nadir} = \alpha \Delta f^*_{ss} = -\frac{\alpha P_d}{D + K_g} \qquad (15)$$

where $\alpha = \Delta f^*_{nadir} / \Delta f^*_{ss}$.

Moreover, the initial RoCoF is calculated as

$$\left.\frac{d\Delta f}{dt}\right|_{t=0} = ab = -\frac{P_d}{2H} \qquad (16)$$

Hence, the coefficient $b$ can be derived as

$$b = \frac{D + K_g}{2\alpha H} \qquad (17)$$

Then the *s*-domain frequency dynamics can be expressed as

$$\Delta f(s) = -\frac{\alpha P_d}{s(2\alpha H s + D + K_g)} \qquad (18)$$

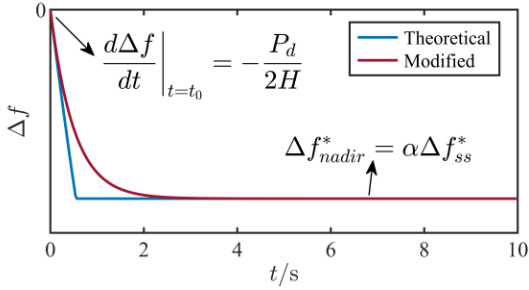

Fig. 4 Theoretical frequency dynamics and its exponential approximation.

Substituting the *s*-domain frequency dynamics into the ASF model shown in Fig. 1, the aggregated optimal AAPC in the feedback form can be derived as follows.

$$G_w(s) = -G_g(s) + K_w \tag{19}$$

where $G_g(s) = \sum_{i \in \mathcal{N}_g} G_{gi}(s)$, $K_w = D - (D+K_g)/\alpha$.

The expression of the optimal AAPC of WTs is free from disturbance $P_d$. Hence, the optimal AAPC for a given power system is independent of the magnitude of the power deficit, which is claimed as event insensitivity. Eq.(19) further indicates that the optimal AAPC of WTs consists of two parts. The part $-G_g(s)$ makes WTs produce a power component opposite to the regulation of the mechanical power of SGs. From the system perspective, this power component simplifies the system frequency model to be first-order, thus eliminating the conventionally convex frequency nadir. On the other hand, according to the features of the governor of SGs, this power component is negative, which conduces to the speed recovery from the WT perspective. Another part $K_w$ produces a power component that increases linearly with the frequency variation, thereby adjusting the value of the frequency nadir.

In the multi-WTs system, the aggregated optimal AAPC needs to be reasonably allocated to each WT, as follows.

$$\begin{cases} G_{wj}(s) = c_j G_w(s), j \in \mathcal{N}_w \\ \sum_{j \in \mathcal{N}_w} c_j = 1 \end{cases} \tag{20}$$

where $c_j$ is the allocation factor of the WT.

In this paper, an allocation strategy considering both the releasable kinetic energy and the increasable active power of the WT is proposed, with details as shown in Appendix C. Therefore, the output power of the WT during the frequency support period of $t \in [t_0, t_f]$ is calculated as

$$P_{ej} = P_{ej0} + G_{wj}(s)\Delta f_j(s), j \in \mathcal{N}_w \tag{21}$$

where $P_{ej0}$ is the steady-state output power of the WT before the power deficit disturbance; $\Delta f_j$ is the local frequency of the WT. It indicates that the proposed optimal AAPC can be deployed locally to WTs like the classic VIC, which shows the good application potential of the optimal AAPC.

Moreover, it should be emphasized that the consistency between the optimal AAPC and the time-domain control relies on the exponential decay form of (14) is a good approximation of the optimal frequency trajectory. That is to say, the shape of the optimal frequency dynamics shown in Fig.3 (c) should be universal for the trajectory optimization model of different power systems and disturbance magnitude. The proof of the required universality is given in Section IV.

### D. Exit Strategy of Frequency Support

Due to the coupling of rotor speed and turbine power, the WT is usually unable to recover to the steady state at the end of the optimal AAPC. In other words, there is a certain deviation between the trajectory of WT derived from the constant turbine power and the actual trajectory. To ensure the state recovery of the WT, the following exit strategy is designed.

For any WT regulated by the optimal AAPC, the following three events will trigger the exit strategy.

$$P_{ej}(t) \leq P_{MPPTj}(t) \text{ or } t \geq t_f \text{ or } \omega_{rj}(t) \leq \omega_r^{\min} \tag{22}$$

where $j \in \mathcal{N}_w$; $P_{MPPTj}$ denotes the MPPT power of the WT; $\omega_{rj}$ denotes the rotor speed of the WT; $\omega_r^{\min}$ denotes the lower limit of rotor speed of the WT.

Denote the activation moment of the exit strategy as $t_{ej}$, the output power of the WT for the exit strategy is designed as

$$P_{ej}(t) = (1-\gamma_j)P_{tj}(t) + \gamma_j P_{MPPTj}(t),\ t \geq t_{ej} \tag{23}$$

where $\gamma_j$ is a coefficient with a value range of $\gamma_j \in [0,1]$.

To eliminate the power saltation caused by control switching, the coefficient $\gamma_j$ is designed as follows: substituting output power of the WT at the $t = t_{ej}$ into (23) to derive the value of $\gamma_j$, if it exceeds the limits, take boundary value.

$$\gamma_j = \frac{P_{ej}(t_{ej}) - P_{tj}(t_{ej})}{P_{MPPTj}(t_{ej}) - P_{tj}(t_{ej})} \tag{24}$$

## IV. UNIVERSALITY OF OPTIMAL FREQUENCY TRAJECTORY

As mentioned above, it is necessary to prove that the shape of the optimal frequency trajectory in Fig. 3(c) is general. Hence, the proof is given as follows.

Integrating the swing equation at the time interval $t \in [t_0, t_f]$ and combining the constraints shown in (6), one obtains

$$\Delta f(t_f) + \frac{D}{2H}S_{\Delta f} = \frac{E_m - P_d t_f}{2H}, \tag{25}$$

where

$$S_{\Delta f} = \int_{t_0}^{t_f} \Delta f(t)dt,\ E_m = \int_{t_0}^{t_f} \Delta P_m(t)dt.$$

$E_m$ is the FR energy of SGs at the time interval $t \in [t_0, t_f]$; $\Delta f(t_f)$ is the terminal frequency at $t = t_f$.

**Theorem 1:** Optimal AAPC of WTs must maximize the FR energy of SGs.

**Proof:** Assuming the FR energy of SGs is not equal under two different AAPC of WTs, it is advisable to set

$$0 < E_{m1} < E_{m2}. \tag{26}$$

In the power deficit cases, both sides of (25) are less than zero, then one obtains

$$\frac{E_{m2} - P_d t_f}{2H} = \kappa \frac{E_{m1} - P_d t_f}{2H} < 0 \tag{27}$$

where $0 < \kappa < 1$.

Denote the frequency at the time interval $t \in [t_0, t_f]$ under $E_{m1}$ and $E_{m2}$ as $\Delta f_1(t)$ and $\Delta f_2(t)$, respectively. One obtains

$$\Delta f_2(t_f) + \frac{D}{2H}S_{\Delta f 2} = \kappa(\Delta f_1(t_f) + \frac{D}{2H}S_{\Delta f 1}) \tag{28}$$

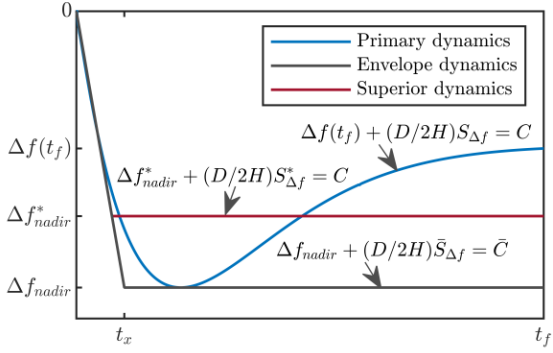

Fig. 5 Frequency trajectory characteristic.

An obvious case satisfying the above equation is
$$\Delta f_2(t) = \kappa \Delta f_1(t), t \in [t_0, t_f] \quad (29)$$
Therefore, the frequency nadir satisfies
$$\Delta f_{nadir2} = \kappa \Delta f_{nadir1} > \Delta f_{nadir1}. \quad (30)$$

The above analysis shows that there certainly exists a higher frequency nadir under the larger FR energy of SGs. Therefore, the optimal AAPC of WTs must maximize the FR energy of SGs, which is an unknown constant under a certain power deficit, denoted as $E_m^{max}$.

**Theorem 2**: The frequency nadir can not be lower than the terminal frequency $\Delta f(t_f)$.

**Proof:** It can be proved by contradiction. Suppose there is a frequency nadir below the terminal frequency, as shown in the blue solid line in Fig. 5.
$$\Delta f_{nadir} < \Delta f(t_f) \quad (31)$$

According to **Theorem 1**, the right-hand side of (25) is a constant with $E_m = E_m^{max}$, denoted as $C$. Therefore, the optimal frequency dynamics satisfy
$$\Delta f(t_f) + (D/2H) S_{\Delta f} = C \quad (32)$$
where $C = (E_m^{max} - P_d t_f)/2H$.

Particularly, the envelope dynamics of the frequency refers that the frequency drops rapidly to the frequency nadir $\Delta f_{nadir}$ with the initial RoCoF and then maintains until terminal time, as shown in the gray solid line in Fig. 5. Correspondingly, the integral of this frequency dynamics is denoted as $\bar{S}_{\Delta f}$. Therefore, for any frequency trajectory with the characteristics shown in (31), one can obtain
$$C = \Delta f(t_f) + (D/2H) S_{\Delta f} > \Delta f_{nadir} + (D/2H) \bar{S}_{\Delta f} = \bar{C}, \quad (33)$$
$$C = \eta \bar{C}, \quad (34)$$
where $0 < \eta < 1$ as $C$ and $\bar{C}$ are both less than 0; the integral of $\bar{S}_{\Delta f} = 0.5(t_f + t_c)\Delta f_{nadir}$, $t_c = t_f - t_x$; $t_x$ is the time to reach the nadir of the envelope frequency trajectory.

For any $\eta \in (0,1)$, the following frequency dynamics with a higher frequency nadir can be easily obtained within the envelope trajectory, which satisfies the constraint of (32). Specifically, the superior frequency dynamics have the same shape as the envelope frequency trajectory, as shown in the red solid line in Fig. 5. Thus, the higher frequency nadir is calculated as

$$\Delta f_{nadir}^* = \mu \Delta f_{nadir} \quad (35)$$

According to (34), the scale factor $\mu$ can be derived as
$$\mu = \frac{2H + Dt_f - \sqrt{\begin{array}{c}4H^2 + D^2 t_f^2 + 4HDt_f^2 - \eta D^2 t_f^2 \\ +\eta D^2 t_c^2 - 4\eta HDt_f + 4\eta HDt_c\end{array}}}{D*(t_f - t_c)} \quad (36)$$

Then, the value range of $\mu$ is derived as $\mu \in (0,1)$
$$\begin{cases} \mu = \dfrac{2H + Dt_f - \sqrt{(2H + Dt_f)^2 - X}}{D*(t_f - t_c)} > 0 \\ \mu = \dfrac{2H + Dt_f - \sqrt{(2H + Dt_e)^2 + Y}}{D*(t_f - t_c)} < 1 \end{cases} \quad (37)$$
where
$X = \eta D^2(t_f^2 - t_c^2) + 4\eta HD(t_f - t_c) > 0$
$Y = (1-\eta)D^2(t_f^2 - t_c^2) + 4(1-\eta)DH(t_f - t_c) > 0$.

Therefore, there is always a higher frequency nadir than the primary nadir if it exceeds the terminal frequency, which indicates that the optimal frequency nadir should not exceed the terminal frequency.

Based on **Theorem 1** and **2**, the frequency nadir maximization is equivalent to minimizing the frequency integral.
$$\max \Delta f_{nadir} \Leftrightarrow \max \Delta f(t_f) \Leftrightarrow \min S_{\Delta f} \quad (38)$$

Obviously, the minimum frequency integral is that the frequency drops rapidly to the frequency nadir with initial RoCoF and then maintains to the terminal time of $t = t_f$, as shown in Fig. 3(c). The above analysis does not presuppose any information about power systems and disturbances, so the conclusion is universal.

## V. CASE STUDY I: A TWO-MACHINE POWER SYSTEM

A two-machine power system consisting of an SG and a WT is first constructed to validate the proposed optimal AAPC, as shown in Fig. 6. The total active load is 150MW. The WT is an aggregate model consisting of $20 \times 5$MW DFIG-based WTs with parameters shown in TABLE I, and the detailed model can refer to [25]. G1 is the steam turbine with a simplified governor model as shown in (39), and the key parameters are listed in TABLE II.
$$G_g(s) = -\frac{K_m(1 + F_H T_R s)}{R(1 + T_R s)} \quad (39)$$

Based on the system parameters and the hypothetic power disturbance of $P_d = 0.1 P_L$, i.e. 15MW, the trajectory optimization model is constructed and solved, deriving the ratio factor of $\alpha = 1.186$. Then the value of $K_w$ in the optimal AAPC of the WF is calculated to be $-14.1$. Combined with $G_g(s)$, the optimal AAPC of the WT is obtained as $G_w(s) = -G_g(s) + K_w$.

### A. Properties of the Optimal AAPC

Based on the above two-machine system, this subsection reveals the essential properties of the optimal AAPC. The wind speed of the WT is 9m/s in this subsection. The power disturbance is set as an active load surge of 15MW, and the output characteristics of the proposed optimal AAPC are shown in Fig. 7. The green solid line in Fig. 7 shows the additional power component produced by the term $-G_g(s)$, which is negative and exactly offsets the increment in the mechanical power of G1. In contrast, the term $K_w$ produces a positive power component to support the system frequency, as shown by the blue solid line in Fig. 7. These two opposite components cause the overall additional power of the WT to increase first for grid frequency support and then decrease for speed recovery, as shown by the red solid line in Fig. 7.

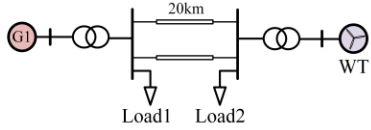

Fig. 6 Two-machine power system.

TABLE I
PARAMETERS OF THE WIND TURBINE

| Symbol | Description | Value |
|---|---|---|
| $S_n$ | Rated capacity (MVA) | 5.556 |
| $P_n$ | Rated power (MW) | 5 |
| $P_e^{max}$ | Maximum power (MW) | 5 |
| $P_e^{min}$ | Minimum power (MW) | 0 |
| $\omega_n$ | Rated rotor speed (rpm) | 12.1 |
| $\omega_r^{min}$ | Minimum rotor speed (p.u.) | 0.7 |
| $J$ | Rotor inertia ($kg \cdot m^2$) | 16,801,544 |

TABLE II
PARAMETERS OF THE SYNCHRONOUS GENERATOR

| Symbol | Description | G1 |
|---|---|---|
| $S_n$ | Rated capacity (MVA) | 200 |
| $K_m$ | mechanical power gain factor | 0.85 |
| $H_g$ | Inertia time constant (s) | 4.2 |
| $R$ | governor regulation | 0.05 |
| $F_H$ | Fraction of high-pressure turbine | 0.3 |
| $T_R$ | Reheat time constant (s) | 8 |

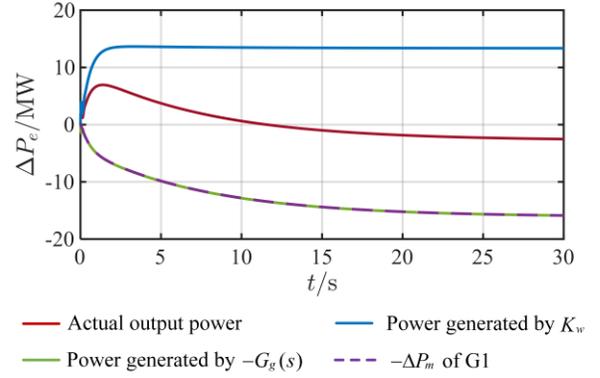

Fig. 7 Active power output characteristics of the optimal AAPC of WTs.

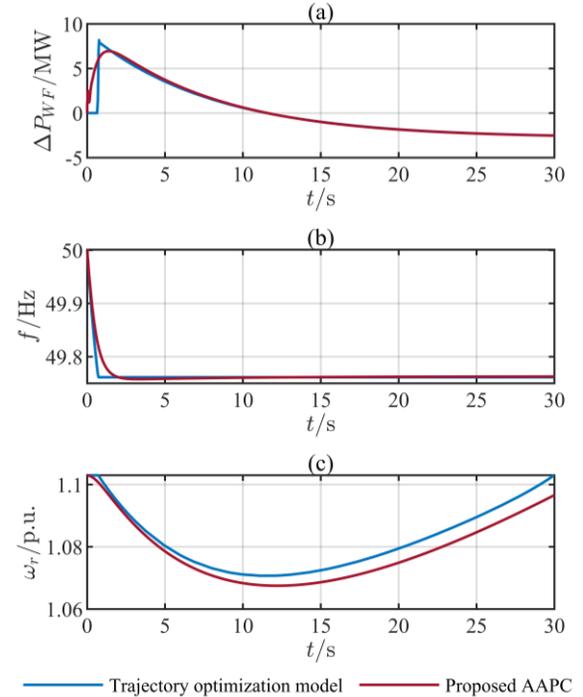

Fig. 8 Comparison of proposed AAPC and trajectory optimization model. (a)Power support of the WF. (b) System frequency. (c) Rotor speed of the WF.

Fig. 8 further compares the theoretically time-domain solution of the trajectory optimization model and the simulation results of the proposed optimal AAPC. As a feedback approximation, the output power dynamic of the proposed AAPC is consistent with the optimal time-domain control curve derived from the trajectory optimization model, as shown in Fig. 8(a). Correspondingly, the system frequency presents the expected first-order shape, and the frequency nadir is the same as that of the trajectory optimization model, as shown in Fig. 8(b). As the proposed AAPC includes a power component that promotes rotor speed recovery, the rotor speed first decreases and then recovers, as shown in Fig. 8(c). However, since the trajectory optimization model ignores wind energy losses, the actual rotor speed does not recover to the steady-state value as expected in theoretical results. This is why a proper exit strategy is needed.

*B. Performance of the Exit Strategy*

The exit strategy aims to ensure the state recovery of the WT without deteriorating the frequency nadir. This subsection verifies the performance of the proposed exit strategy, and the MPPT-based exit strategy in [16] is used as a comparison.

Admittedly, the MPPT-based exit strategy eliminates power saltation if the output power curve intersects the MPPT curve. However, this case is not inevitable. Fig. 9 and Fig. 10 show the comparison between the proposed exit strategy and the MPPT-based exit strategy in the above two scenarios. Fig. 9 indicates that both the proposed exit strategy and MPPT-based method can smoothly switch to the MPPT mode for speed recovery when the output power curve intersects with the MPPT curve. However, when the output power curve does not intersect with the MPPT curve, the MPPT-based exit strategy will cause the output power of the WT to plummet to the MPPT curve at the moment of activation, as shown by the blue dotted line in Fig. 10(a). As a result, this power plummet causes a secondary frequency drop, as shown by the blue dotted line in Fig. 10(b). In contrast, the proposed exit strategy maintains the smoothness of the output power while achieving speed recovery as shown by the orange solid line in Fig. 10. To conclude, the proposed exit strategy is effective and ensures the stable operation of the WF without deteriorating the frequency nadir.

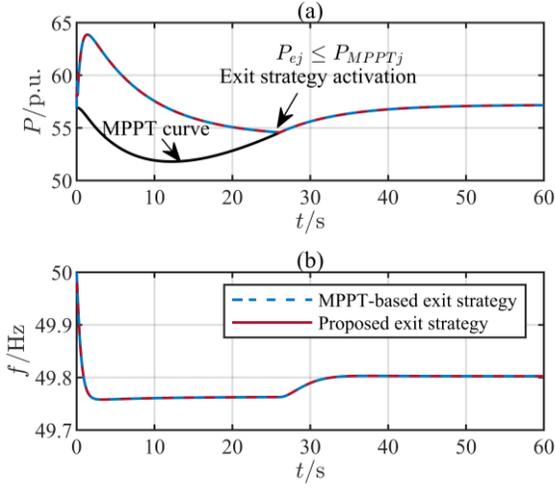

Fig. 9 Validation of exit strategy with intersection point. (a) Output power of the WT. (b) Grid frequency dynamics.

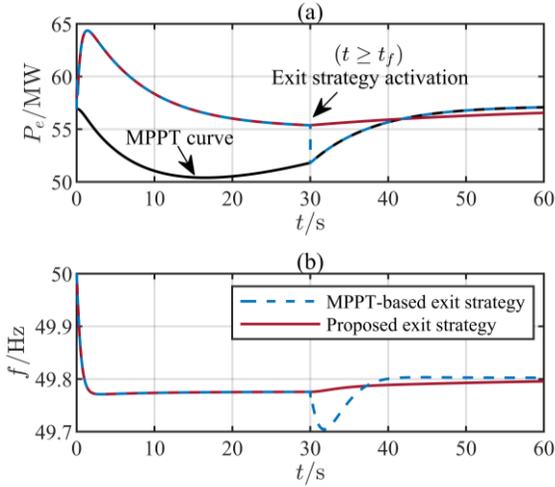

Fig. 10 Validation of exit strategy without intersection point. (a) Output power of the WT. (b) Grid frequency dynamics.

*C. Event Insensitivity*

As analyzed in Section III.C, the optimal AAPC for a given power system is theoretically free from the magnitude of the power deficit, namely event insensitivity. Fig. 11 shows the frequency dynamics under various power disturbances, and the optimal AAPC is derived from a hypothetic power deficit of $P_d = 0.1P_L$. The referred frequency nadir can be solved based on the trajectory optimization model, denoted as $\Delta f_{nadir}^*$. The results indicate that the optimal AAPC makes the grid frequency exhibit the desired first-order dynamics regardless of the disturbance size. The convex frequency nadir is eliminated and the frequency nadir can be improved as the ideal solution of the trajectory optimization model.

Considering the operating constraints of the actual WT, the event-insensitive range of the optimal AAPC should be limited. Taking the frequency nadir as the indicator, Fig. 12 further tests the optimal AAPC under a series of power deficits. The results show that within a certain range of power disturbance, the optimal AAPC can improve the frequency nadir as expected. However, if the FR demand under the extremely severe power deficit scenarios that exceed the frequency support capability of the WT, the performance of the optimal AAPC will inevitably deviate from the reference. Despite performance degradation, the comparison in Fig. 12 indicates that the optimal AAPC is still significantly better than the classic VIC under severe scenarios. Therefore, the optimal AAPC of the WT is event-insensitive over a relatively wide range of scenarios.

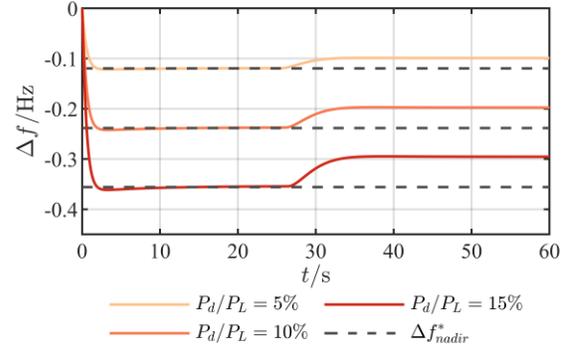

Fig. 11 Frequency dynamics in the event-insensitive range of optimal AAPC.

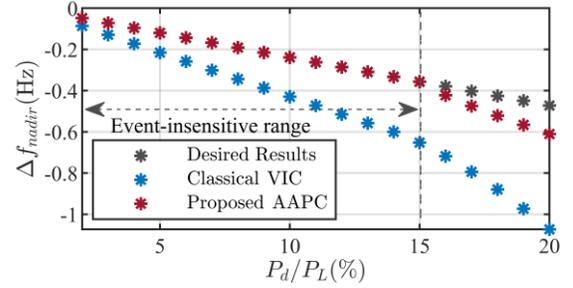

Fig. 12 Validation of the event insensitivity of the optimal AAPC.

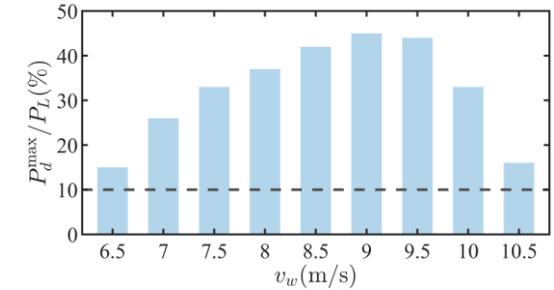

Fig. 13 Impact of wind speed of the WT on the event-insensitive range of the optimal AAPC.

To quantitate the performance degradation of the optimal AAPC, the indicator is defined as follows.

$$e_r = \frac{\Delta f_{nadir} - \Delta f_{nadir}^*}{\Delta f_{nadir}^*} \times \% \qquad (40)$$

In this paper, the event-insensitive range is adopted as $e_r \leq 5\%$. Under this criterion, Fig. 13 shows the impact of the wind speed of the WT on the event-insensitive range of the optimal AAPC, where $P_d^{max}$ denotes the upper limit of the event-insensitive range. The results indicate that the optimal AAPC of WTs is event-insensitive to a general major-power disturbance such as a 10% load surge at different wind speeds.

The above results indicate that the optimal AAPC derived from a specified disturbance is also optimal for diverse disturbance magnitudes. Hence, the optimal AAPC of WTs can be solved and deployed based on the on-line rolling update under a hypothetical disturbance, avoiding the heavy post-event computational burden.

## VI. CASE STUDY II: IEEE 39-BUS SYSTEM

The IEEE 39-bus system shown in Fig. 14 is constructed to verify the effectiveness of the proposed AAPC in the large-scale power system. Specifically, equivalent SG of the external system G1 and steam turbines G2/G5/G6/G7/G8/G9 are equipped with IEEEG1 type of governor; hydro turbines G3/G10 are equipped with HYGOV type of governor; gas turbine G4 is equipped with GAST type of governor. The detailed model of the above governors can be found in [26]. In addition, 5 aggregate WTs are connected to the system at Bus 8/14/16/18/22, denoted as WT1, WT2, WT3, WT4, and WT5 respectively.

Each aggregated WT is composed of $80 \times 5MW$ DFIG-based WTs, and the wind speed is set to 6.5m/s in WT1, 7.5m/s in WT2, 8.5m/s in WT3, and 9.5m/s in WT4, and 10.5m/s in WT5, to model the diversity of operating conditions of WTs. The model and parameters of the WT are the same as in Section V.

First, the trajectory optimization model is constructed and solved, and the aggregated optimal AAPC of all the WTs can be obtained, which needs to be reasonably allocated to each WT. Correspondingly, the allocation factors of WTs are derived as $c_1 = 0.08$, $c_2 = 0.21$, $c_3 = 0.374$, $c_4 = 0.248$, $c_5 = 0.0875$.

The performance of the proposed optimal AAPC is compared with the following three inertial control strategies.

*Strategy 1*: Classical VIC in [2], that is, the coefficients of the proportional-derivative (PD) controller remain constant. In this paper, $k_f = 20$, and $k_{in} = 10$.

*Strategy 2*: Adaptive VIC in [3]. Specifically, the coefficients of VIC are adaptively adjusted according to the releasable kinetic energy of the WT.

*Strategy 3*: Optimal temporary frequency support control (TFSC) in [16]. The differential controller in the traditional VIC is replaced with a first-order low-pass filter.

Then, the following two typical scenarios of power shortage are designed to verify the performance of the above methods.

*Scenario 1*: Active load of the system surges by 10%.
*Scenario 2*: G7 trips.

For a large-scale power system, the grid frequency inevitably appears spatial and temporal distribution differences post a major power deficit, taking into account the oscillation of SGs. Given this, a widely used concept is introduced in this paper, namely the frequency of the center of inertia (CoI), which can reflect the average frequency dynamics of the grid. According to [27], the frequency of the CoI is calculated as follows.

$$f_{CoI} = \frac{\sum_{i \in \mathcal{N}_g} H_{gi} S_{gi} f_i}{\sum_{i \in \mathcal{N}_g} H_{gi} S_{gi}} \quad (41)$$

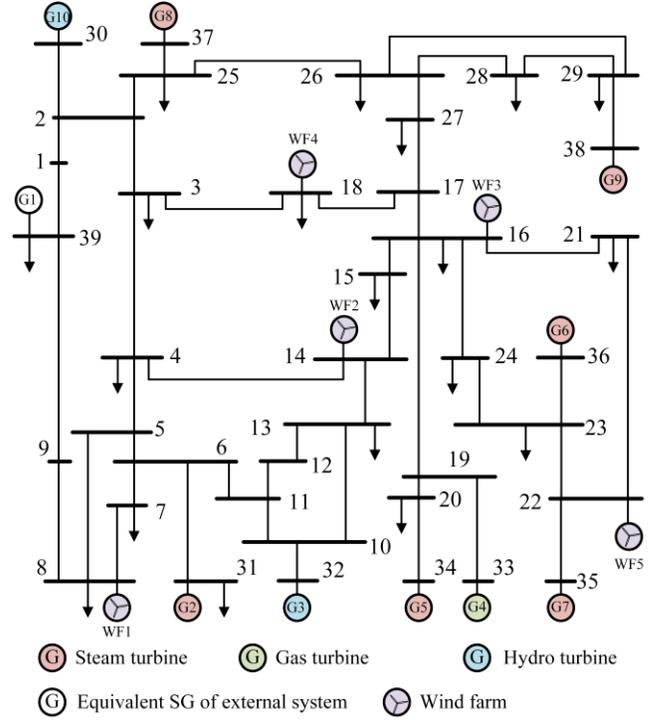

Fig. 14 Modified IEEE 39-bus system.

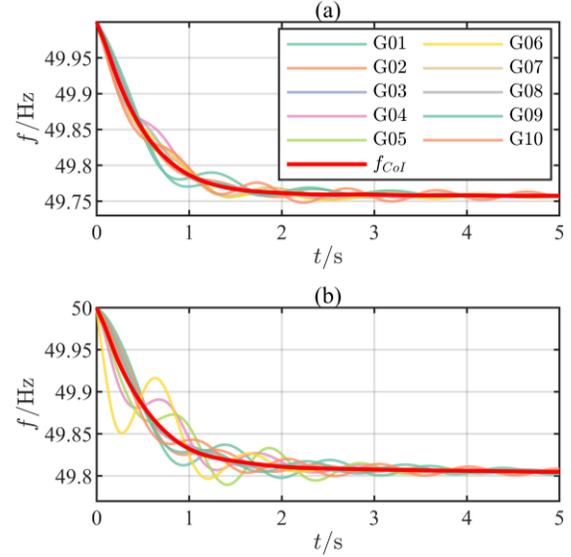

Fig. 15 Frequency dynamics of SGs and CoI. (a) Scenario 1. (b) Scenario 2.

Fig. 15 shows the frequency dynamics of each SGs and the CoI in **scenario 1** and **scenario 2**, with WTs equipped with the optimal AAPC. It can be seen that the frequency of each SGs shows certain oscillation characteristics at the initial stage of a power disturbance. Fortunately, the frequency of the CoI eliminates the frequency difference caused by the oscillatory behavior of SGs and properly depicts the average grid frequency dynamics, as shown in the red solid line in Fig. 15. Hence, it is reasonable to model the overall grid dynamics by the frequency of the CoI. If there is no special description, the frequency in the following refers to that of the CoI.

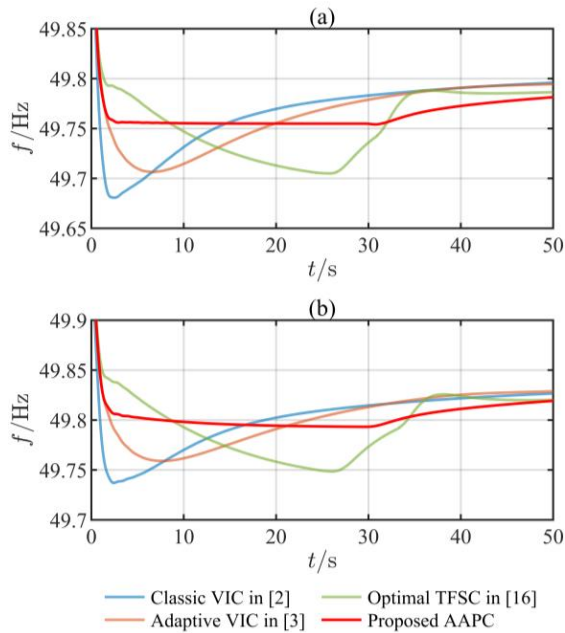

Fig. 16 Comparison of system frequency dynamics under different control strategies. (a) Load surge. (b) Generator tripping.

TABLE III
COMPARISON OF FREQUENCY NADIR IN SCENARIO 1&2

|  | $\Delta f_{nadir}$ (Hz) | |
| --- | --- | --- |
|  | Scenario 1 | Scenario 2 |
| No FR | -0.4133 | -0.3304 |
| Strategy 1 | -0.3193 | -0.2630 |
| Strategy 2 | -0.2936 | -0.2410 |
| Strategy 3 | -0.2950 | -0.2515 |
| **Proposed AAPC** | **-0.2458** | **-0.2068** |

Moreover, it should be noted that the frequency of the CoI is only used for result display. The dependent frequency of the implementation of the optimal AAPC is still measured locally.

The grid frequency dynamics with the proposed control strategy and Strategies 1-3 in Scenarios 1&2 are compared in Fig. 16, and the deviation of frequency nadirs in the transient process are listed in TABLE III, where "No FR" denotes that WTs do not provide frequency support. The results in TABLE III indicate that the temporary FR of WTs does significantly improve the frequency nadir of the power system post a major power deficit. Taking **Scenario 1** as an example, the frequency nadir deviation with WTs under the above control strategies is improved by 22.74% (**Strategy 1**), 28.96% (**Strategy 2**), 28.62% (**Strategy 3**), 40.53% (**optimal AAPC**) respectively, compared with "No FR". Furthermore, the comparison results indicate that the proposed optimal AAPC is effective in a large-scale power system with multiple WTs, which contributes to a higher frequency nadir compared with the existing control strategies. Specifically, compared with strategies 1-3, the optimal AAPC of WTs improves the frequency nadir deviation in the above two scenarios by 23.02% and 21.37% (**Strategy 1**), 16.28% and 14.19% (**Strategy 2**), 16.68% and 17.78% (**Strategy 3**), respectively.

To demonstrate the effectiveness of the proposed allocation strategy among WTs, it is compared with the average allocation principle. The dynamics of the system frequency and WTs with the proposed allocation strategy and average allocation in Scenario 1 are compared as shown in Fig. 17. The results show that the proposed allocation strategy takes into account the differences in operating conditions and power support margins of WTs. The reasonable allocation factor effectively avoids the over-lower-limit of the rotor speed of the WT with low wind speed (WT1), as shown in Fig. 17(b); as well as avoids the over-upper limits of the output power of the WT with high wind speed (WT5), as shown in Fig. 17(a). Over-lower limits may cause concerns about the safe operation of the WT. As for the over-upper-limit of the power reference, although the limiter ensures a bearable actual output power of the WF, the deviation of the output power from the expected value will inevitably result in a degradation in control performance, i.e. deteriorating the frequency nadir, as shown in Fig. 17(c). The above results show that the proposed allocation strategy among WTs improves the transient frequency stability of the power system while ensuring the safe operation of the WT.

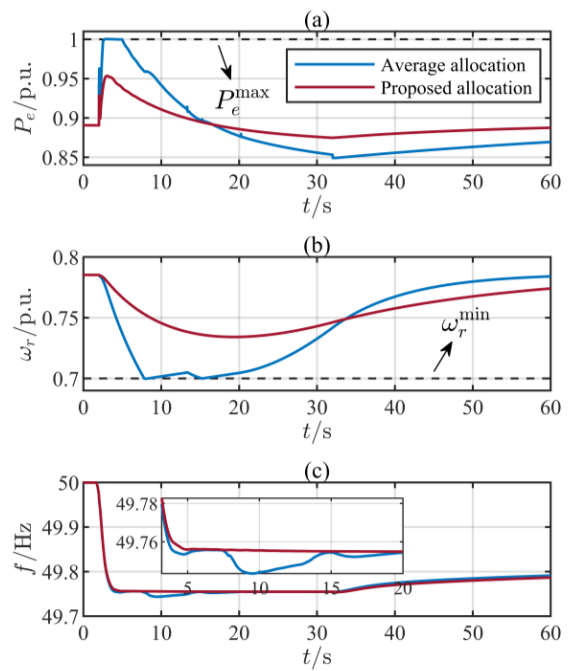

Fig. 17 Dynamics of WTs under different allocation strategies. (a) Output power of WT5 under high wind speed. (b) Rotor speed of WT1 under low wind speed. (c) System frequency.

## VII. CONCLUSION

In this paper, a system-view optimal AAPC of WTs is proposed to provide fast frequency support for power systems. Different from the widely used VIC methods, this paper explores the optimal frequency support mode of WTs based on the system demands. As a feedback approximation, the proposed AAPC of WTs inherits the optimal performance on maximizing the frequency nadir of the trajectory optimization model. More importantly, the universality of the reverse approximation is strictly proved. Besides, the event insensitivity of the proposed method avoids the dependence on the disturbance magnitude. Hence, the optimization calculation can be preposed by rolling implementation to avoid the heavy post-event communication and computation burden. Extensive simulation results prove that the proposed method can effectively improve the system frequency stability post a major power deficit.

## APPENDIX A
### DYNAMIC OF THE GOVERNOR SYSTEM

In the ASF model, the dynamics of separate governor is retained, and the FR can be expressed as

$$\Delta P_{mi}(s) = G_{gi}(s)\Delta f(s) \quad (42)$$

where $i \in \mathcal{N}_g$.

Ignoring the nonlinearity of governors, the following state-space form of FR of governors can be derived from (42).

$$\begin{cases} \Delta \dot{x}_{gi}(t) = A_{gi}\Delta x_{gi}(t) + B_{gi}\Delta f(t) \\ \Delta P_{mi}(t) = C_{gi}\Delta x_{gi}(t) + D_{gi}\Delta f(t) \end{cases} \quad (43)$$

where $i \in \mathcal{N}_g$; $x_{gi} \in \mathbb{R}^{m_i \times 1}$ are the state variables of the $i$-th governor; $A_{gi} \in \mathbb{R}^{m_i \times m_i}$, $B_{gi} \in \mathbb{R}^{m_i \times 1}$, $C_{gi} \in \mathbb{R}^{1 \times m_i}$, and $D_{gi} \in \mathbb{R}$ are the coefficient matrices; $m_i$ is the orders of the governor.

Then, the overall dynamics of the governor system can be expressed as follows.

$$\begin{cases} \Delta \dot{x}_g(t) = A_g \Delta x_g(t) + B_g \Delta f(t) \\ \Delta P_g(t) = C_g \Delta x_g(t) + D_g \Delta f(t) \end{cases} \quad (44)$$

where

$$x_g = \begin{bmatrix} x_{g1} \\ x_{g2} \\ \vdots \\ x_{gM} \end{bmatrix}, A_g = \begin{bmatrix} A_{g1} & & & \\ & A_{g2} & & \\ & & \ddots & \\ & & & A_{gM} \end{bmatrix}, B_g = \begin{bmatrix} B_{g1} \\ B_{g2} \\ \vdots \\ B_{gM} \end{bmatrix},$$

$$C_g = \begin{bmatrix} C_{g1} & C_{g2} & \cdots & C_{gM} \end{bmatrix}, D_g = \sum_{i \in \mathcal{N}_g} D_{gi}$$

## APPENDIX B
### PRINCIPLE OF THE GAUSS PSEUDOSPECTRAL METHOD

In the Gauss pseudospectral method, the interpolation points are discretely distributed in the time interval of $\tau \in [-1,1]$. Hence, the time interval needs to be converted into

$$\tau = \frac{2t}{t_f - t_0} - \frac{t_f + t_0}{t_f - t_0} \quad (45)$$

Taking the roots of the K-order Legendre polynomial as the discrete points, denoted as $\mathcal{K} = \{\tau_1 \ \tau_2 \ \cdots \ \tau_K\}$, the K-order Lagrangian interpolation of the state variable is as follows.

$$x(\tau) \approx X(\tau) = \sum_{i=0}^{K} L_i(\tau) x(\tau_i) \quad (46)$$

where $L_i(\tau) = \prod_{j=0, j \neq i}^{K} \frac{\tau - \tau_j}{\tau_i - \tau_j}$ is the basic function of Lagrangian interpolation.

Similarly, the control variables should also be discretized.

$$u(\tau) \approx U(\tau) = \sum_{i=0}^{K} L_i(\tau) u(\tau_i) \quad (47)$$

Notably, the endpoint of the interval $\tau_f = -1$ is not involved in the above interpolation points, whereas constraints related to the final value of the states are common. Therefore, an estimation for $X_f$ needs to be constructed by the Gauss integration algorithm. First, the theoretical final value of the state can be derived from the dynamics of the system.

$$x(\tau_f) = x(\tau_0) + \int_{-1}^{1} f(x(\tau), u(\tau), \tau) d\tau \quad (48)$$

Then, its discrete estimation can be calculated by

$$X(\tau_f) = X(\tau_0) + \frac{\tau_f - \tau_0}{2} \sum_{k=1}^{K} \omega_k f(X(\tau_k), U(\tau_k), \tau; t_0, t_f) \quad (49)$$

where $\omega_k$ is Gaussian weight.

Then, the differential of the state variable is approximated by interpolation, as follows.

$$\dot{x}(\tau_k) \approx \dot{X}(\tau_k) = \sum_{i=0}^{K} \dot{L}_i(\tau_k) X(\tau_i) = \sum_{i=0}^{K} D_{ki} X(\tau_i) \quad (50)$$

where $D \in \mathbb{R}^{K \times (K+1)}$ is the differential matrix formed by the differential of the basic functions of Lagrangian interpolation.

Finally, the differential equations of the dynamic system can be converted to the following algebraic equations.

$$\sum_{i=0}^{K} D_{ki} X(\tau_i) - \frac{t_f - t_0}{2} f(X(\tau_k), U(\tau_k), \tau; t_0, t_f) = 0 \quad (51)$$

The continuous path constraints are also discretized into the following form.

$$C(X(\tau_k), U(\tau_k), \tau_k; t_0, t_f) \leq 0 \quad (52)$$

So far, the infinite-dimensional continuous trajectory optimization problem has been transformed into a nonlinear programming (NLP) problem, for which there are many efficient solvers, such as the sequential quadratic programming (SQP) method [28].

## APPENDIX C
### ALLOCATION STRATEGY OF MULTI-WTS

To quantify the frequency support capability of the WT, the following two indices are introduced, denoting its maximum releasable kinetic energy and the maximum increasable active power.

$$\begin{cases} \Delta E_{kj}^{\max} = E_{kj0} - E_{kj}^{\min} \\ \Delta P_{ej}^{\max} = P_{ej}^{\max} - P_{ej0} \end{cases}, j \in \mathcal{N}_w \quad (53)$$

where $E_k = 0.5 J \omega_r^2$ is the kinetic energy of the WT. The subscript "0", "min", and "max" denote the initial value, minimum value, and maximum value, respectively.

Larger $\Delta E_{kj}^{\max}$ and $\Delta P_{ej}^{\max}$ indicates a stronger frequency support capacity, but they are irreconcilable, as shown in Fig. 5. It can be seen that sufficient kinetic energy means that the WT operates at a high wind speed, whereas the increasable power is limited for the large steady-state power, and vice versa. Therefore, the allocation factor of WTs is designed as

$$c_j^* = \min\left\{ \frac{\Delta E_{kj}^{\max}}{\sum_{j \in \mathcal{N}_w} \Delta E_{kj}^{\max}}, \frac{\Delta P_{ej}^{\max}}{\sum_{j \in \mathcal{N}_w} \Delta P_{ej}^{\max}} \right\}, j \in \mathcal{N}_w \quad (54a)$$

$$c_j = \frac{c_j^*}{\sum_{j \in \mathcal{N}_w} c_j^*}, j \in \mathcal{N}_w \quad (54b)$$

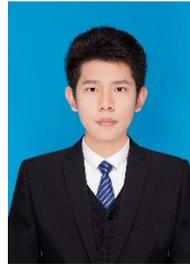

**Yubo Zhang.** (S'21) received the B.S. degree in electrical engineering from Xi'an Jiaotong University, China, in 2019. He is currently pursuing the Ph.D. degree in the school of Electrical Engineering in Xi'an Jiaotong University. His main field of interest includes the control for renewable energy, and power system frequency stability.

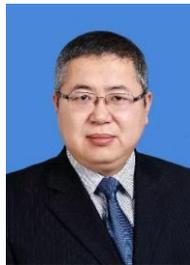

**Zhiguo Hao.** (M'10) was born in Ordos, China, in 1976. He received his B.Sc. and Ph.D. degrees in electrical engineering from Xi'an Jiaotong University, Xi'an, China, in 1998 and 2007, respectively. He has been a Professor with the Electrical Engineering Department, Xi'an Jiaotong University. His research interest includes power system protection and control.

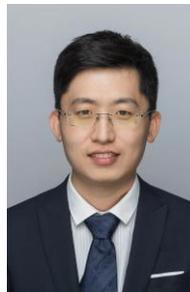

**Songhao Yang.** (S'18-M'19) was born in Shandong, China, in 1989. He received the B.S. and Ph.D. degrees in electrical engineering from the Xi'an Jiaotong University, Xi'an, China, in 2012 and 2019, respectively. Besides, he received the Ph.D. degree in electrical and electronic engineering from Tokushima University, Japan, in 2019.

Currently, he is an Associate Professor at Xi'an Jiaotong University. His research interest includes power system control and protection.

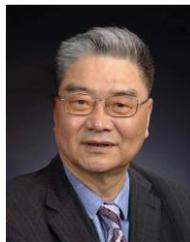

**Baohui Zhang.** (SM'99-'F'19) was born in Hebei Province, China, in 1953. He received the M.Eng. and Ph.D. degrees in electrical engineering from Xi'an Jiaotong University, Xi'an, China, in 1982 and 1988, respectively. He has been a Professor in the Electrical Engineering Department at Xi'an Jiaotong University since 1992. His research interests are system analysis, control, communication, and protection.